\documentstyle[12pt]{article}

\begin{document}

\title{Perturbative study of multiphoton processes in the tunneling regime}

\author{Marco Frasca\footnote{e-mail: marcofrasca@mclink.it} \\
Via Erasmo Gattamelata, 3 \\
00176 Roma (Italy)}

\maketitle

\abstract{
A perturbative study of the Schr\"{o}dinger
equation in a strong electromagnetic field with dipole approximation
is accomplished in the Kramers-Henneberger frame.
A prove that just
odd harmonics appear in the spectrum for a linear polarized
laser field is given, assuming that the
atomic radius is much lesser than the free-electron quiver motion
amplitude. Within this approximation
a perturbation series is obtained in the Keldysh
parameter giving a description of multiphoton processes in the
tunneling regime. The theory is applied to
the case of hydrogen-like atoms: The
spectrum of higher order harmonics and the
above-threshold ionization rate are derived.
The ionization rate computed in this way determines
the amplitudes of the harmonics.
The wave function of the atom proves to be rigid with respect
to the perturbation so that the effect of the laser field on the
Coulomb potential in the computation of the probability amplitudes
can be neglected as a first approximation: This
approximation improves as the ratio between the amplitude
of the quiver motion of the electron and the atom radius becomes larger.
The semiclassical description currently adopted for harmonic generation
is so rederived by solving perturbatively the Schr\"{o}dinger
equation.
}

\newpage

Availability of powerful sources of laser light has permitted,
in recent years, the realization of experiments through
gaseous media that have shown several new physical effects as
photoionization with a number of photons absorbed by the
electron well above the ionization threshold and generation
of a broad range of harmonics of the laser frequency \cite{kei}. This
latter effect could have a lot of technological applications
and, as such, has been widely studied both theoretically
and experimentally.

The possibility to turn a physical effect into a practical
application is strongly linked with the availability of
a satisfactory theoretical model. But, it is common belief
that, due to the intensity of the laser field, no
perturbation theory can be done. The main aim
of this paper is then to show how perturbation theory
can be straightforwardly applied also for intense
laser fields and analytical expressions can be
computed for any kind of multiphoton process, at least
for hydrogen-like atoms. The development parameter turns
out to be the the square root of the ratio between the
ionization energy $I_B$ and the ponderomotive energy
$U_p$ proportional to the intensity of the laser field,
known in literature as the Keldysh parameter $\gamma$.
The regime of a small Keldysh parameter characterize the
so-called tunnelling regime that is the one of interest here.

Theoretical approaches to multiphoton processes are
non-perturbative in nature and
resort to Floquet theory as in \cite{potv}, numerical
methods applied directly to the Schr\"{o}dinger equation
as done firstly in \cite{ebe} or semiclassical models
\cite{cork1}. On the basis of the semiclassical
ideas, a quantum theory for harmonic generation
has been obtained by L'Huillier and coworkers in \cite{cork2}:
Our theory permits to justify the main assumptions of the
quantum theory of these authors, so that, in turn,
the semiclassical ideas prove
to be a fairly good description of harmonic generation.

The approach we apply to the Schr\"{o}dinger equation for
an atom in an electromagnetic field can be easily understood
using a two-level model, widely used for harmonic generation
\cite{twol}. This model has the Hamiltonian (here and in the
following we will take $\hbar=c=1$)
\begin{equation}
    H=\frac{\omega_0}{2}\sigma_3+\Omega\cos(\omega t)\sigma_1
\end{equation}
being $\omega_0$ the level separation, $\Omega$ the intensity
of the laser field and $\omega$ its frequency, $\sigma_1$ and
$\sigma_3$ are Pauli matrices. If $\Omega$ is small with
respect to $\omega_0$, standard perturbation theory applies
by interaction picture through an unitary transformation that
removes the unperturbed part of the Hamiltonian: This gives
a Dyson series in the small development parameter
$\Omega/(\omega_0\pm\omega)$, out of resonance.
Recently, duality has been introduced in
perturbation theory \cite{fra1} and a dual interaction picture
has been devised where one does an unitary transformation to
remove the perturbation. For the above Hamiltonian one has to
take $U=e^{-i\sigma_1\frac{\Omega}{\omega_0}\sin(\omega t)}$ that
yields the transformed hamiltonian \cite{fra11}
\begin{eqnarray}
    H_F&=&\frac{\omega_0}{2}e^{2i\sigma_1\frac{\Omega}{\omega_0}
    \sin(\omega t)}\sigma_3 \nonumber \\
    &=&\frac{\omega_0}{2}
    J_0\left(\frac{2\Omega}{\omega}\right)\sigma_3+
    \frac{\omega_0}{2}\sum_{n\neq 0}
    J_n\left(\frac{2\Omega}{\omega}\right)e^{in\sigma_1\omega t}
    \sigma_3
\end{eqnarray}
where now, perturbation theory can be done for
$\Omega\gg\omega_0,\omega$. We see straightforwardly that the
unperturbed part of the Hamiltonian is ``dressed'' by the
laser field and so, the energy levels are shifted. Then,
the perturbation has odd and even harmonics of the
laser frequency and both can appear in the spectrum.
But, probability amplitudes that enters in the computation
of the spectrum do not depend on the unitary transformations
one does on the Hamiltonian and the states. So, we have
sketched the physics of the two-level model in an intense
monochromatic field just through dual interaction picture. Although,
as we will show, the two-level model does not apply for
current experiments with atomic samples
as in this case one observes just odd harmonics
in agreement with our full theory and it is not just
a problem of a proper experimental setup,
nevertheless it could have a wide range of applications in magnetic
resonance experiments, for some other kind of media
as optical cavities \cite{mey} or wherever the
conditions one meets for atomic samples are no more fulfilled.

The dual interaction picture applies in the same way
also to the Schr\"{o}dinger equation in
a semiclassical laser field and in the dipole
approximation, as currently treated in literature \cite{cork2}.
The correspondence with the two-level model above
is remarkable. The Hamiltonian in this case is
\begin{equation}
    H=\frac{{\bf p}^2}{2m}+V({\bf x})+\frac{e}{m}{\bf A}(t){\bf p}
    +\frac{e^2}{2m}{\bf A}^2(t).
\end{equation}
By the unitary transformation
$U(t)=\exp\left(-i\frac{e}{m}\int_0^tdt'{\bf A}(t'){\bf p}
-i\frac{e^2}{2m}\int_0^tdt'{\bf A}^2(t')\right)$
the above Hamiltonian transforms into
\begin{equation}
    H_{KH}=U^{\dagger}(t)\left(\frac{{\bf p}^2}{2m}+V({\bf x})\right)U(t)
    =\frac{{\bf p}^2}{2m}+V[{\bf x}+{\bf a}(t)]
\end{equation}
being ${\bf a}(t)=-\frac{e}{m}\int_0^tdt'{\bf A}(t')$. This is
the well-known Kramers-Henneberger Hamiltonian
and the unitary transformation above define the
so called Kramers-Henneberger frame \cite{KH}
that shows as the effect of the electromagnetic field is
to introduce a time-dependent translation on the potential of
the unperturbed Hamiltonian by a length ${\bf a}(t)$.
The laser field can be modeled as
${\bf A}(t)=-\frac{E(t)}{\omega\sqrt{1+\xi^2}}
[\hat{\bf x}\cos(\omega t)+\xi\hat{\bf y}\sin(\omega t)]$
for a general ellipticity parameter $\xi$.
Here we consider the simplest case of a linear polarization
$\xi=0$ and an instant rising of the laser field, that is
$E(t)=const$. So, one has \cite{kei}
\begin{equation}
    H_{KH}=\frac{{\bf p}^2}{2m}+
    \int_{-\lambda_L}^{\lambda_L}\frac{dx'}{\pi}
    \frac{V(x-x',y,z)}{\sqrt{\lambda_L^2-x'^2}}+
    \sum_{k=1}^{+\infty}i^k[e^{ik\omega t}+(-1)^ke^{-ik\omega t}]
    v_k({\bf x})
\end{equation}
with
\begin{equation}
    v_k({\bf x})=
    \int_{-\lambda_L}^{\lambda_L}\frac{dx'}{\pi}
    V(x-x',y,z)\frac{T_k\left(\frac{x'}{\lambda_L}\right)}
    {\sqrt{\lambda_L^2-x'^2}} \label{eq:vk}
\end{equation}
being $T_k(x)=\cos(k\arccos(x))$ the $k$-th Chebyshev polynomial
of first kind and
$\lambda_L=\frac{eE}{m\omega^2}=
\sqrt{\frac{4U_p}{m}}\frac{1}{\omega}$ the maximum free-electron
quiver motion excursion.
This length is pivotal in the study of atoms in an
intense laser field as generally
one has $\lambda_L\gg a$, being $a=\frac{1}{mZe^2}$
the Bohr radius. One can see that,
as for the two-level model, we have the potential of the
unperturbed part of the Hamiltonian ``dressed'' by the laser field
and all the harmonics, odd and even, are present in the perturbation.
We can now show that, in all the current experiments where the potential
$V({\bf x})$ depends just on $r=|{\bf x}|$ and $\lambda_L\gg a$, being
$a$ the Bohr radius of the atoms in the sample, then just odd harmonics
appear in the spectrum. Indeed, we can rewrite eq.(\ref{eq:vk}) as
\begin{equation}
    v_k({\bf x})=\int_{-1}^1dx'V(\sqrt{(x-\lambda_L x')^2+y^2+z^2})
    \frac{T_k(x')}{\pi\sqrt{1-x'^2}}. \label{eq:vkr}
\end{equation}
If the laser field is enough intense, a series in $\frac{a}{\lambda_L}$
is obtained if one develops eq.(\ref{eq:vkr}) in Taylor series as
\begin{eqnarray}
    v_k({\bf x})&=&\int_{-1}^1dx'
    \left.V(\sqrt{(x-\lambda_L x')^2+y^2+z^2})\right|_{x=0,y=0,z=0}
    \frac{T_k(x')}{\pi\sqrt{1-x'^2}} \nonumber \\
    & &-x\int_{-1}^1dx'
    V'(\lambda_L |x'|)\frac{x'}{|x'|}
    \frac{T_k(x')}{\pi\sqrt{1-x'^2}}+\cdots.
\end{eqnarray}
Despite its appearance, the terms of this series can be evaluated
for a Coulomb potential and proved to be finite assuring the convergence.
This is due to the fact that in this case the integrals can be
computed analitically. Then,
from the above expression two main conclusions can be drawn. Firstly,
multiphoton effects are due to a dipole induced on the atom
in the same direction as the electric field of the laser and
secondly, Chebyshev polynomials have a definite parity and due to the
symmetrical range of integration, only odd polynomials give a non-null
contribution to the second term, while the first term has no physical
consequences and in the following will be neglected. So, only odd
harmonics contribute to the spectrum while, even harmonics are 
quadrupole radiation and then strongly depressed.
Indeed, for a Coulomb potential one obtains
\begin{equation}
    v_{2n+1}({\bf x})\approx -i(-1)^n\frac{x}{\lambda_L}(2n+1)
    \frac{Ze^2}{\lambda_L}.        \label{eq:hl}
\end{equation}
This result, that does not involve any other approximation beside the
simmetry of the potential and the amplitude of the quiver motion of the
electron with respect to the atomic radius, supports in some way the physical view
recently given in \cite{kklp}, where it is assumed that the electron
recolliding with the atomic core, emits bremsstrahlung radiation
that is cut off at the maximum amplitude of the quiver motion of the
electron, producing in this way just odd harmonics. 

To complete the above discussion before introducing perturbation theory, we
have to study the ``dressed'' potential $v_0$.
This should be managed differently
from the time-dependent part. Indeed, we have to separate the original
potential $V(r)$ from the shifts induced by the laser field on
the energy levels of the atom. This can be obtained by a Taylor
expansion as
\begin{eqnarray}
    v_0({\bf x})&=&\int_{-1}^1\frac{dx'}{\pi}
    \frac{V(\sqrt{(x-\lambda_L x')^2+y^2+z^2})}{\sqrt{1-x'^2}}
    \nonumber \\
    &=&V(r)+\delta_LV({\bf x}) \nonumber \\
    &=&V(r)+
    \frac{\lambda_L^2}{4r^3}\left[V'(r)y^2+V'(r)z^2+V''(r)x^2r\right]
    +\cdots
\end{eqnarray}
where is seen that only even terms survive and higher order terms
fall off very rapidly with $r$. The above expression
assumes a very simple form for a Coulomb potential
\begin{equation}
    v_0({\bf x})=-\frac{Ze^2}{r}\left[1+
    \sum_{n=1}^{+\infty}A_n\left(\frac{\lambda_L}{r}\right)^{2n}
    P_{2n}\left(\frac{x}{r}\right)\right] \label{eq:dp}
\end{equation}
being $A_n=\int_{-1}^1dxx^{2n}/(\pi\sqrt{1-x^2})$ and $P_n$ the $n$-th
Legendre polynomial. This way to express the dressed Coulomb potential
gives us a way to prove that the wave function is ``rigid'' with
respect to the perturbation using standard Rayleigh-Schr\"{o}dinger
perturbation scheme, for the kind of problems we discuss here. But,
it should be pointed out that for stabilization things are quite
different \cite{fed}.

The equations for the amplitudes are given by
\begin{eqnarray}
    i\dot{a}_m(t)&=&
                  \sum_{n\neq m}a_n(t)
                  \langle m|\delta_L V({\bf x})|n\rangle
                  e^{-i(\tilde{E}_n-\tilde{E}_m)t}+
                  \nonumber \\
                 & &\sum_n\sum_{k=1}^{+\infty}i^k a_n(t)
                  \langle m|v_k({\bf x})|n\rangle
                 [e^{-i(\tilde{E}_n-\tilde{E}_m-k\omega)t}+
                  (-1)^ke^{-i(\tilde{E}_n-\tilde{E}_m+k\omega)t}]
                  \label{eq:amp}
\end{eqnarray}
having set $\tilde{E}_n=E_n+\langle n|\delta_L V({\bf x})|n\rangle$,
being $\delta_L V({\bf x})$ the part of the static potential due to
the laser field. At this point, all the machinery of standard
perturbation theory applies \cite{mes}. For our aim, we have to
show that the Rayleigh-Schr\"{o}dinger part gives indeed a small
contribution to the amplitudes. By assuming the atom initially in
its ground state, this contribution is
\begin{equation}
    a_m^{RS}(t)\approx\frac{
                  \langle m|\delta_L V({\bf x})|1\rangle}
                  {\tilde{E}_1-\tilde{E}_m}
                  (e^{-i(\tilde{E}_1-\tilde{E}_m)t}-1).
                  \label{eq:corr}
\end{equation}
Using eq.(\ref{eq:dp}) is easy to verify that no contribution comes
for $m=2$ as $\langle 2|\delta_L V({\bf x})|1\rangle=0$ but the
degeneracy of level 2 is removed by the dressed potential as one has
$\langle m=2,l=1,l_z=0|\delta_L V({\bf x})|m=2,l=1,l_z=0\rangle
=Ze^2/(240a)(\lambda_L/a)^2$ and
$\langle m=2,l=1,l_z=\pm 1|\delta_L V({\bf x})|m=2,l=1,l_z=\pm 1\rangle
=-Ze^2/(480a)(\lambda_L/a)^2$ while
$\langle m=2,l=0,l_z=0|\delta_L V({\bf x})|m=2,l=0,l_z=0\rangle=0$.
Indeed, one can see that all the states
having $m$ even do not give a first order contribution even if the
level shift is not null, while the level-shift is always $0$
when $l=0$. Instead, for $m=3$ one has e.g.
$\langle m=3,l=2,l_z=0|\delta_L V({\bf x})|m=1,l=0,l_z=0\rangle
=Ze^2\sqrt{150}/(10800a)(\lambda_L/a)^2$ and for the level shifts
$\langle m=3,l=2,l_z=0|\delta_L V({\bf x})|m=3,l=2,l_z=0\rangle
=Ze^2/(5670a)(\lambda_L/a)^2-Ze^2/(136080a)(\lambda_L/a)^4$
and $\langle 1|\delta_L V({\bf x})|1\rangle=0$, so
the correction of eq.(\ref{eq:corr}) turns out to be
\begin{equation}
    a_{3,2,0}^{RS}(t)\approx
                  -\frac{
                  \frac{\sqrt{150}}{10800}
                  \left(\frac{\lambda_L}{a}\right)^2
                  }
                  {\frac{4}{9}+
                  \frac{1}{5760}\left(\frac{\lambda_L}{a}\right)^2
                  -\frac{1}{136080}
                  \left(\frac{\lambda_L}{a}\right)^4
                  }
                  (e^{i\frac{8}{9}E_1 t}-1)
\end{equation}
that is indeed negligeable and the wave function turns out to be
``rigid'' with respect to the deformations
introduced by the laser field. This is even more true as larger become the
ratio $\lambda_L/a$. The reason for this is that only a finite number
of terms of eq.(\ref{eq:dp}) give a non-null contribution to the
matrix elements. It is interesting to note that
for stabilization of an atom
in intense laser field the situation is exactly the contrary as one
should be able to diagonalize the Hamiltonian
$H_0={\bf p}^2/2m+v_0({\bf x})$ being
the time-dependent part negligible,
an approximation that becomes exact in the limit of
infinite frequency of the laser field \cite{kei,fed}.

Then, the iterative procedure
to solve eq.(\ref{eq:amp}) can be applied to compute the probability
transition for any process. This approach implies that off-resonant
contributions should be sistematically neglected. In this way,
a golden rule is straightforwardly obtained as
\begin{eqnarray}
    P_{i\rightarrow f}=2\pi\sum_{n=1}^{+\infty}
    |\langle i|v_n({\bf x})|f\rangle|^2
    \delta\left[\tilde{E}_f-\tilde{E}_i-n\omega\right] \label{eq:gr}
\end{eqnarray}
from which several results for multiphoton processes can be obtained.
It is assumed a
continuum of final states to sum over so that excited levels can decay,
otherwise quantum resonance theory applies
\cite{rg} and Rabi flopping is obtained. In any case, going to
second order gives a.c.Stark shifts of the energy levels.
Rabi frequency due to resonance with the $k$-th harmonic of the
perturbation with two levels $m$ and $n$ of the atom is (ref.\cite{rg})
$\frac{\Omega_R}{2}=|\langle m|v_k({\bf x})|n\rangle|$.

From eq.(\ref{eq:gr}) we can easily compute the rate of above threshold
ionization. For hydrogen-like atoms [eq.(\ref{eq:hl})] and assuming the
atom initially in its ground state one has
\begin{equation}
    \Gamma=\frac{32}{3}\frac{\omega^2}{U_p}\gamma^2
    \sum_{n=n_0}^{+\infty}
    \left[\frac{I_B}{(2n+1)\omega}\right]^\frac{5}{2}
    \left[1-\frac{I_B}{(2n+1)\omega}\right]^\frac{3}{2} \label{eq:Gamma}
\end{equation}
being $n_0$ the minimun integer for which $(2n_0+1)\omega-I_B\ge 0$.
It has been used the fact that, as shown above,
for the ground state of hydrogen-like
atoms there is no shift by the part of the static potential due to the
laser field, that is $\langle 1|\delta_L V({\bf x})|1\rangle=0$ for Coulomb
potential. Beside, a plane wave is assumed for the particle in the final
state to make computation simpler. By taking ref. \cite{hui} for
experimental results , we can check the above expression
for helium and neon that show a large plateau in the tunneling regime.  
So, we have $U_p=$ 155 eV being the intensity
1.5 $\times$ $10^{15}$ W/cm$^2$, $\omega=$ 1.177 eV and
$I_B=$ 24.59 eV. Then, $\gamma\approx$ .4
and $\Gamma\approx$ 0.026 eV, that is small as it should be
expected. The same computation for neon gives approximatively 0.02 eV.

To analyse the question of harmonic generation, one has to compute
$<x>=\langle\Psi(t)|x|\Psi(t)\rangle$. To complete this computation,
we assume that no intermediate resonance is present and will
justify this assumption a posteriori through the quantum resonance
theory of ref.\cite{rg}, that here applies. So, let us take an atom
initially prepared in its ground state as to have
$a_i(0)=\delta_{i0}$. From eq.(\ref{eq:amp}) one has
\begin{eqnarray}
    a_m(t)&=& \delta_{m0}
                  +\frac{\langle m|\delta_L V({\bf x})|0\rangle}
                  {\tilde{E}_0-\tilde{E}_m-i\epsilon}
                  e^{-i(\tilde{E}_0-\tilde{E}_m-i\epsilon)t}+
                  \nonumber \\
                 & &\sum_{k=1}^{+\infty}i^k a_n(t)
                  \langle m|v_k({\bf x})|0\rangle\left[
                 \frac{e^{-i(\tilde{E}_0-\tilde{E}_m-k\omega-i\epsilon)t}}
                 {\tilde{E}_0-\tilde{E}_m-k\omega-i\epsilon}+
                 (-1)^k\frac{e^{-i(\tilde{E}_0-\tilde{E}_m+k\omega-i\epsilon)t}}
                  {\tilde{E}_0-\tilde{E}_m+k\omega-i\epsilon}\right]+\cdots
\end{eqnarray}
with the limit $\epsilon\rightarrow 0$ understood as to have
$\frac{1}{x\pm i0}=P\frac{1}{x}\mp i\pi\delta(x)$, being $P$ the
principal value.
As is customary in perturbation theory, we keep just those terms that are
near resonant with the harmonics of the perturbation: The only possibility
left is the continuous spectrum, as it should be with the current
understanding of harmonic generation. So, we take
\begin{equation}
    a_{\bf p}(t)\approx
                 -\sum_{k=1}^{+\infty}i^k
                  \langle {\bf p}|v_k({\bf x})|0\rangle
        (-1)^k\frac{e^{i(E_{\bf p}-\tilde{E}_0-k\omega+i\epsilon)t}}
                  {E_{\bf p}-\tilde{E}_0-k\omega+i\epsilon} \label{eq:ap}
\end{equation}
being $\bf p$ the momentum of the particle in the continuous part of the
spectrum. Now, we specialise this expression to the case of hydrogen-like
atoms having
\begin{equation}
    a_{\bf p}(t)\approx\frac{Ze^2}{\lambda_L^2}
                 \sum_{n=0}^{+\infty}(2n+1)
               \langle{\bf p}|x|0\rangle
            \frac{e^{i(E_{\bf p}-E_0-(2n+1)\omega+i\epsilon)t}}
                  {E_{\bf p}-E_0-(2n+1)\omega+i\epsilon}.
\end{equation}
Then, for the dipole moment one has
\begin{equation}
    <x>\approx\sum_{\bf p}a_{\bf p}(t)e^{-i(E_{\bf p}-E_0)t}
    \langle 0|x|{\bf p}\rangle+c.c.
\end{equation}
After passing from the sum to integration through
$\sum_{\bf p}\rightarrow V\int\frac{d^3p}{(2\pi)^3}$
and taking for the final state a plane wave, one
gets the final expression for the harmonic spectrum
\begin{equation}
    <x>\approx -\frac{64}{3^\frac{9}{2}}\frac{Ze^2\omega}{U_p^2}
    \gamma^5
    \sum_{n=n_0}^{+\infty}
    \frac{x_n^\frac{3}{2}}{\left(x_n+\frac{\gamma^2}{3}\right)^5}
    \sin((2n+1)\omega t)
\end{equation}
being $x_n=\frac{(2n+1)\omega-I_B}{3U_p}$. The normalization to
$3U_p$ for $x_n$ originates from the fact that, from the above expression,
the intensities of the harmonics
reduce as the factor $3U_p$ increases. Then, if the
Keldysh parameter $\gamma$ is enough small we can take
\begin{equation}
    <x>\approx -\frac{64}{3^\frac{9}{2}}\frac{Ze^2\omega}{U_p^2}
    \gamma^5
    \sum_{n=n_0}^{+\infty}
    \frac{1}{x_n^{\frac{7}{2}}}
    \sin((2n+1)\omega t)
\end{equation}
so that, only for $x_n\leq 1$ the harmonic amplitudes are large.
This is the approximate cut-off law found out through semiclassical
methods in ref.\cite{cork1}. It should also be stressed the
existence of a minimun harmonic order $n_0$ that should be expected
due to the close connection between harmonic generation and multiphoton
ionization. Indeed, this lower bound comes out from the phase space
through the integration of the Dirac function both for the golden rule
(\ref{eq:gr}) and for the computation of the dipole moment $<x>$.  
Then, one gets $n_0=10$ and $9$ for helium 
and neon respectively, that means harmonic $21$ for the starting
point of the spectrum in the regime of interest.
It should be pointed out that the above equation for $<x>$ has to take
properly into account the ionization rate $\Gamma$ of eq.(\ref{eq:Gamma})
as to have at last
\begin{equation}
    <x>\approx -\frac{64}{3^\frac{9}{2}}\frac{Ze^2\omega}{U_p^2}
    \gamma^5
    \sum_{n=n_0}^{+\infty}
    \frac{x_n^\frac{3}{2}}{\left(x_n+\frac{\gamma^2}{3}\right)^5}
    \sin((2n+1)\omega t)e^{-\Gamma t}.
\end{equation}
One can estimate the constant factor that determines the amplitude
of the harmonics $\frac{64}{3^\frac{9}{2}}\frac{Ze^2\omega}{U_p^2}\gamma^5$.
Indeed, for helium one obtains  approximately .32 $\times$ 10$^{-8}$ eV$^{-1}$ and
for neon about .12 $\times$ 10$^{-7}$ eV$^{-1}$, showing, as it should be,
a larger amplitude for neon.

A further analysis concerns the effect of intermediate resonances on
the spectrum of harmonics. On the basis of the theory of ref.\cite{rg},
one can write down eq.(\ref{eq:ap}) as
\begin{equation}
    a_{\bf p}(t)\approx
                 -\sum_{k=1}^{+\infty}i^k
                  \langle {\bf p}|v_k(r)|0\rangle
        (-1)^k\frac{e^{i(E_{\bf p}-\tilde{E}_0-k\omega+i\epsilon)t}}
                  {E_{\bf p}-\tilde{E}_0-k\omega+i\epsilon}
                  \cos\left(\frac{\Omega_R}{2}t\right)
\end{equation}
with $\Omega_R$ the Rabi frequency computed taking in account the
resonances between the ground state and other discrete levels. To
compute the above expression we assumed that the atom is initially
prepared in its ground state so that,
$a_0(t)=\cos\left(\frac{\Omega_R}{2}t\right)$, essentially the
rotating wave approximation. It
is easy to realize that one gets the harmonics in the spectrum
shifted by the quantity $\pm\frac{\Omega_R}{2}$.

The theory above could have wide applicability as,
in principle for any multiphoton
process one is able to compute analytical formulae to compare with
experimental results. For instance,
an improvement easy to implement is to use a full
Coulomb wave function also for the final state in the above computations.
On the other hand, even if major features of
multiphoton processes are described by this theory, several problems
are surely opened up as the applicability of the theory for
an ellipticity parameter $\xi\neq 0$, the introduction of a slower
rising of the laser field or how to take into account
all the features that real experiments have for harmonic generation.
Beside, when the intensity of the laser field becomes too high the
above approach should be properly modified
as relativistic effects enter into the physical
picture and, e.g. even harmonics can also be significant \cite{kk}.
Experiments to generate even harmonics are also carried out
using solid surfaces as in \cite{ago}.
Anyhow, it should be stressed how the possibility to derive a
perturbative solution to the Schr\"{o}dinger equation could give a
chance to check models of multiphoton physics that no other
approach offers.

\end{document}